\documentclass[aps,prl,amsmath,amssymb,reprint,twocolumn,superscriptaddress,showpacs]{revtex4-1}
\usepackage{epsfig,amssymb}
\usepackage{graphicx}
\usepackage{dcolumn}
\usepackage{bm}
\usepackage{dcolumn}
\usepackage{multirow}
\begin{document}

\title{Satellite Band Structure in Silicon Caused by Electron-Plasmon Coupling}

\author{Johannes~Lischner}
\email{jlischner597@gmail.com}
\affiliation{Department of Physics, University of California,
  Berkeley, California 94720, USA, and Materials Sciences Division,
  Lawrence Berkeley National Laboratory, Berkeley 94720, USA.}
\author{G. K. P{\'a}lsson}
\affiliation{Department of Physics, University of California,
  Davis, California 95616, USA, and Materials Sciences Division,
  Lawrence Berkeley National Laboratory, Berkeley 94720, USA.}
\author{Derek Vigil-Fowler}
\affiliation{Department of Physics, University of California,
  Berkeley, California 94720, USA, and Materials Sciences Division,
  Lawrence Berkeley National Laboratory, Berkeley 94720, USA.}
\author{S. Nemsak}
\affiliation{Department of Physics, University of California,
  Davis, California 95616, USA, and Materials Sciences Division,
  Lawrence Berkeley National Laboratory, Berkeley 94720, USA.}
\author{J. Avila}
\affiliation{Synchrotron SOLEIL, Saint Aubin, BP 48 91192 Gif-sur-Yvette, France}
\author{M. C. Asensio}
\affiliation{Synchrotron SOLEIL, Saint Aubin, BP 48 91192 Gif-sur-Yvette, France}
\author{C. S. Fadley}
\affiliation{Department of Physics, University of California,
  Davis, California 95616, USA, and Materials Sciences Division,
  Lawrence Berkeley National Laboratory, Berkeley 94720, USA.}
\author{Steven G. Louie}
\affiliation{Department of Physics, University of California,
  Berkeley, California 94720, USA, and Materials Sciences Division,
  Lawrence Berkeley National Laboratory, Berkeley 94720, USA.}

\begin{abstract}
  We report the first angle-resolved photoemission measurement of the
  wave-vector dependent plasmon satellite structure of a
  three-dimensional solid, crystalline silicon. In sharp contrast to
  nanomaterials, which typically exhibit strongly wave-vector
  dependent, low-energy plasmons, the large plasmon energy of silicon
  facilitates the search for a plasmaron state consisting of
  resonantly bound holes and plasmons and its distinction from a weakly interacting
  plasmon-hole pair. Employing a first-principles theory, which is
  based on a cumulant expansion of the one-electron Green's function
  and contains significant electron correlation effects, we obtain
  good agreement with the measured photoemission spectrum for the
  wave-vector dependent dispersion of the satellite feature, but without
  observing the existence of plasmarons in the calculations.
\end{abstract}

\maketitle

\emph{Introduction.---} Within the contemporary view of condensed
matter physics\cite{louie2006conceptual} in the Fermi liquid paradigm,
the electronic structure of materials is described in terms of a
quasiparticle picture, where particle-like excitations (such as those
measured in transport or photoemission experiments) in an otherwise
strongly interacting electron system are characterized by weakly
interacting quasi-electrons and quasi-holes, consisting of the bare
particles and a surrounding “screening cloud” of electron-hole pairs
and collective excitations. One example of such collective excitations
are plasmons, quantized charge density oscillations resulting from the
long-range nature of the Coulomb interaction. Both the energy and the
dispersion relation of plasmons depend sensitively on the
dimensionality of the material. In three-dimensional materials, the
energy required to excite a plasmon is typically multiple electron
volts, but in two- and one-dimensional systems, such as doped
graphene\cite{ju2011graphene} or metallic carbon
nanotubes\cite{lin1997low}, plasmons can be gapless excitations with
strong wave-vector dependence and vanishing energy in the zero
wave-vector limit.

The interaction with plasmons has an important effect on the
properties of electrons and holes in solids. For example, the energy
dispersion relation of the electrons in a crystal (the band structure)
is modified. As a more drastic consequence of strong electron-plasmon
coupling, Lundqvist\cite{hedin1967new} predicted the emergence of a
new kind of composite quasiparticles, called plasmarons \footnote{To
  avoid confusion, we point out that the term ``plasmaron'' has also
  been used to describe coupled plasmon-phonon states, see for example
H. Yu and J. C. Hermanson, Phys. Rev. B 40, 11851 (1989).}, consisting of
resonantly bound plasmons and holes, which give rise to additional
sharp features from the conventional quasiparticle peaks, known as the
satellite structures, in photoemission and tunneling spectra. Recent
experiments on doped
graphene\cite{Rotenberg2,walter2011effective,brar2010observation} and
two-dimensional electron gases in semiconductor quantum
wells\cite{Dial} observed prominent satellite structures, which were
interpreted as signatures of plasmaron excitations.

\begin{figure*}
  \includegraphics[width=14cm]{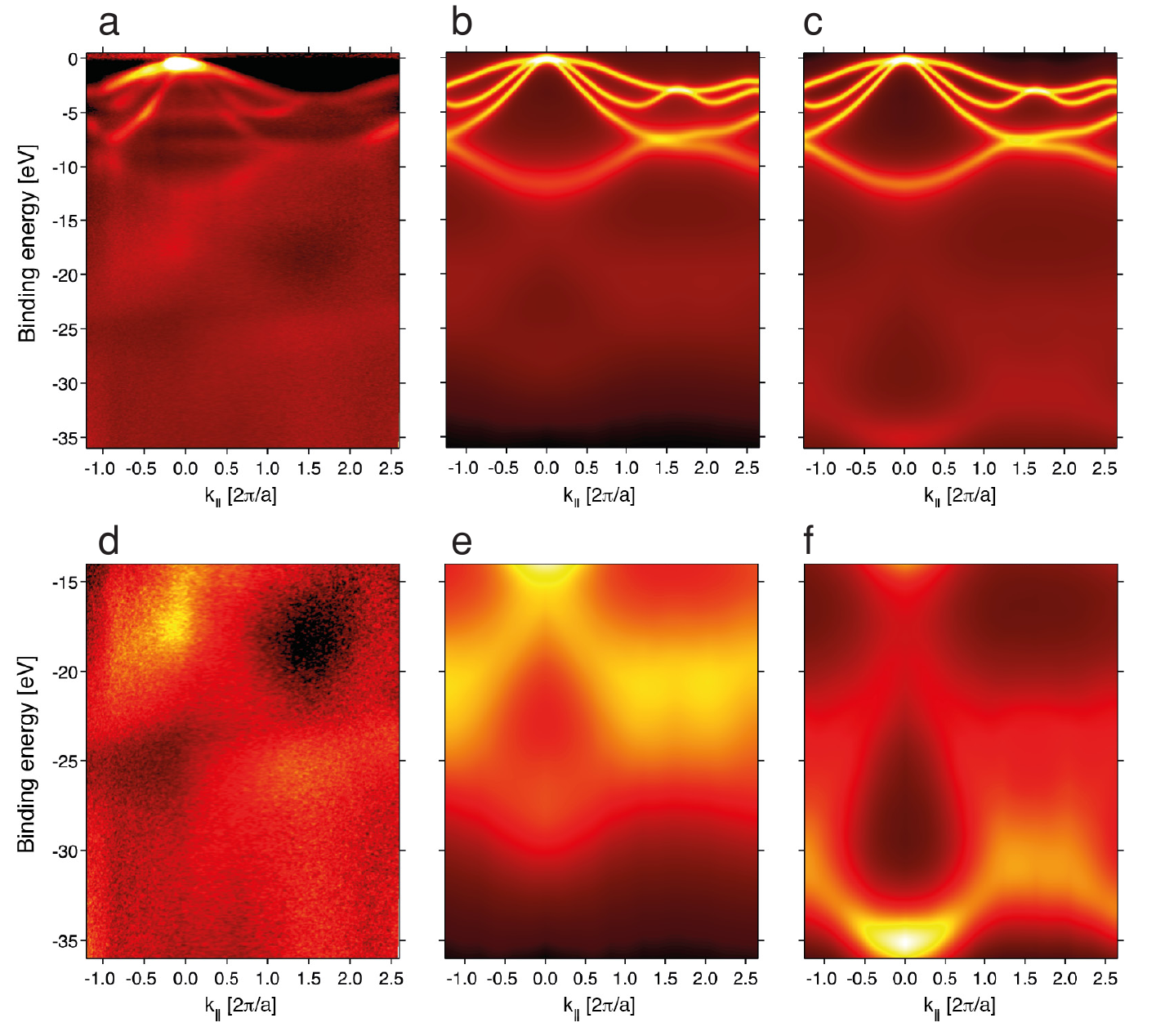}
  \caption{(a): Experimental photoemission spectrum of silicon taken
    along $\phi=-30^{\circ}$ (see appendix for
    definition of $\phi$) using a photon energy of 711 eV. Here,
    $k_{||}$ denotes the component of electron wave vector parallel to
    the surface. (b) and (c): Theoretical photoemission spectra from
    GW plus cumulant theory and GW theory, respectively along
    $\phi=-30^{\circ}$. (d), (e) and (f): Same spectra as in (a), (b)
    and (c), but only the binding energy range relevant to the first
    satellite feature is shown.}
  \label{fig:fig1}
\end{figure*}

Other studies\cite{lischner2013physical,lischner2014satellite} pointed
out that the observed satellite features could also result from the
creation of weakly interacting plasmon-hole pairs instead of strongly
interacting plasmaron states. Such “shake-up” satellites are well
known in the photoemission spectroscopy of molecules, where they
result from the creation of an electron-hole pair or a vibrational
mode in addition to the quasi-hole in the photo-excitation
process. Because of the low plasmon energy in two-dimensional systems
(which is proportional to the square root of the plasmon wave vector)
and other experimental complications, such as the dielectric screening
from a substrate, it has been difficult to identify unambiguously from
experiment whether the observed satellites originate from plasmarons
or shake-up processes involving plasmons.

\begin{figure*}
  \includegraphics[width=14cm]{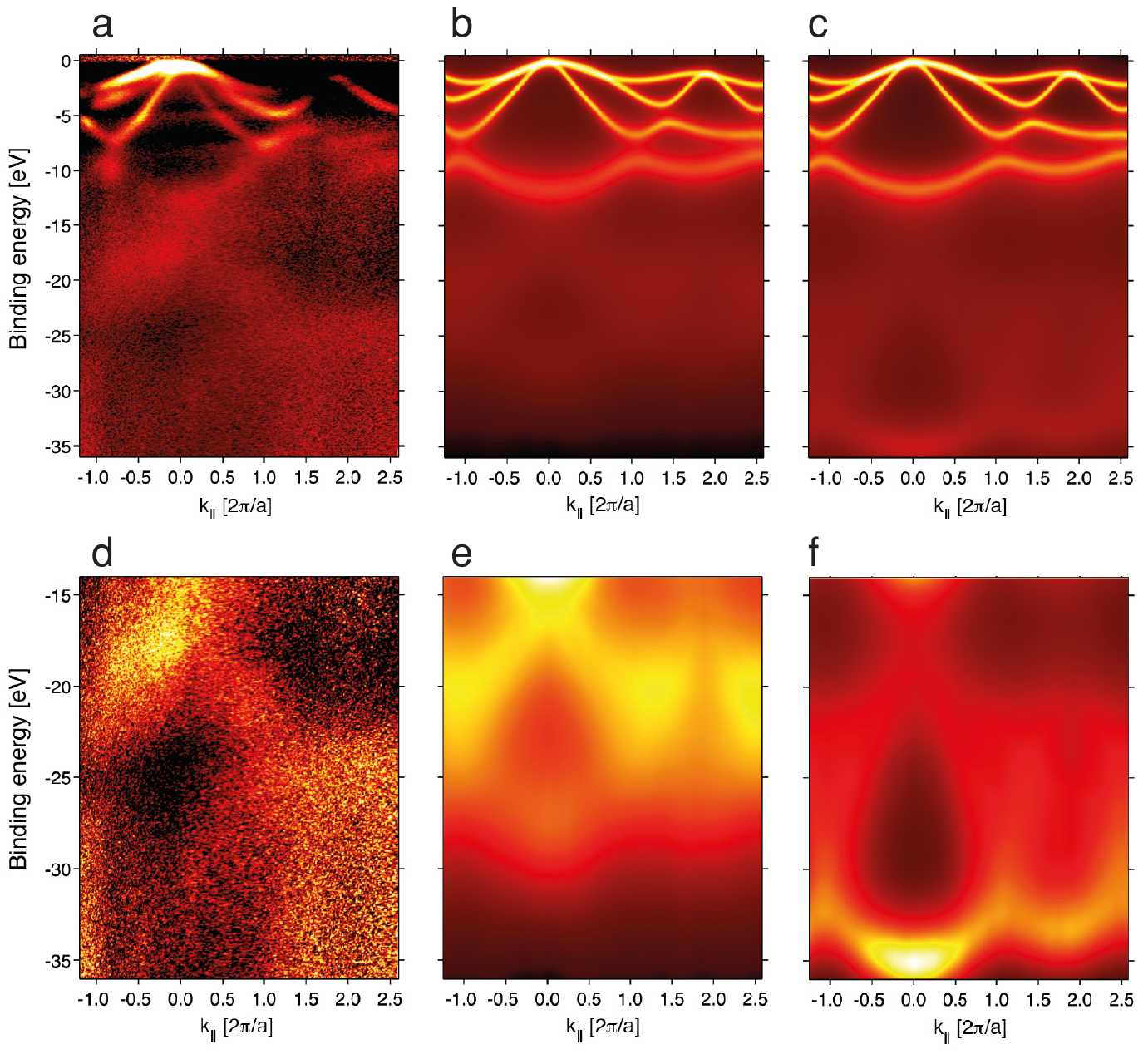}
  \caption{(a): Experimental photoemission spectrum along
    $\phi=-60^{\circ}$ (see appendix for a description of the
    experimental photoemission setup). (b) and (c): Theoretical
    photoemission spectra from GW plus cumulant theory and GW theory,
    respectively. (d), (e) and (f): Same spectra as in (a), (b) and
    (c), but only the binding energy range relevant to the first
    satellite is shown.}
  \label{fig:fig2}
\end{figure*}

In three-dimensional systems, the plasmon energy is much larger than
in two- and one-dimensional systems (it approaches a large constant
value at small wave vectors plus a term which is proportional to the
square of the plasmon wave vector) resulting in significant energy
differences between possible plasmaron states and unbound hole-plasmon
pairs. Also, possible complications from environmental screening are
eliminated. However, obtaining angle-resolved photoemission spectra of
bulk satellite features requires higher energy photons because of the
higher binding energy of the satellites and also the need to minimize
surface related effects. So far, satellite properties in
three-dimensional solids were only probed in angle-integrated
photoemission experiments\cite{Guzzo,ley1972x}, but such experiments
do not give direct insights into satellite properties associated with
individual quasiparticle states, such as their line widths and
dispersions.

To elucidate the nature of the plasmon satellites in three-dimensional
solids, we chose silicon as a prototypical system. It is one of the
most studied and technologically important three-dimensional
semiconductor materials, and a full understanding of its electronic
structure including the wave-vector dependent satellite properties is
highly desirable. Accurate knowledge of the electron-plasmon and
light-plasmon interactions is particularly important for current and
future plasmonic devices\cite{walters2010silicon,gabrielli2009silicon,pillai2007surface}.

\emph{Results---.} Figure 1(a) shows the measured angle-resolved
photoemission spectrum from the [111] surface of silicon along the
$\phi=-30^{\circ}$ direction (see appendix) using photons with an
energy of 711 eV. The spectrum exhibits prominent sharp, dispersive
features at binding energies smaller than 13 eV corresponding to the
usual quasiparticle excitations (i.e., the band states). At binding
energies higher than 15 eV, we observe a more diffuse satellite band
structure, which looks like a fainter, broadened copy of the
quasiparticle band structure. Figure 2(a) shows the measured
angle-resolved photoemission spectrum along the $\phi=-60^{\circ}$
direction and exhibits similar features to the spectrum obtained along
$\phi=-30^{\circ}$.

To gain insight into the observed photoemission spectra, we compare
them to state-of-the-art theories of electronic excitations in
condensed matter systems. Such theories yield spectral functions,
$A_{n\bm{k}}(\omega) = 1/\pi \times |\text{Im}G_{n\bm{k}}(\omega)|$,
which are proportional to the angle-resolved photoemission spectrum
within the sudden approximation\cite{damascelli}. Here, $n$ and
$\bm{k}$ are the band index and the wave vector of the hole created in
the photoemission process, respectively, and $G_{n\bm{k}}(\omega)$
denotes the wave vector and frequency-dependent interacting
one-particle Green's function. Calculations of the Green's function
typically proceed by evaluating a set of Feynman diagrams, which
represent interaction processes between the electrons and other
excitations\cite{mattuck2012guide}.

The GW method\cite{HedinBook,LouieHybertsen} has been used to analyze
photoemission experiments and, recently, to interpret satellite
features for two-dimensional
systems\cite{Rotenberg2,Dial,walter2011effective}. This approach
captures the complicated, dynamic polarization response of the
electron sea to the appearance of a hole in the photoemission process
by approximating the electron self-energy as the first term in a
Feynman series expansion in the screened Coulomb interaction, but it
neglects the contribution of other higher order Feynman diagrams
describing additional correlation effects between electrons. For
low-energy quasiparticle properties, such as the electronic band gaps
and quasiparticle dispersion relations of semiconductors and
insulators, the GW approach has resulted in very good agreement with
experimental measurements from first
principles\cite{LouieHybertsen}. However, much less is known about its
accuracy for satellite properties. For the special case of a
dispersionless hole (such as the hole resulting from the removal of an
electron from a tightly bound atomic core state) interacting with
plasmons, the GW approach fails dramatically to describe the satellite
properties\cite{Guzzo,lischner2013physical,HedinAryasetiawan}. The
exact solution of this model problem can be obtained using a cumulant
expansion of the Green's function\cite{Langreth}. The resulting
spectral function exhibits an infinite series of satellite peaks,
separated by the plasmon energy from the quasiparticle peak and from
each other. The GW approach instead predicts a single satellite peak
separated from the quasiparticle peak by 1.5 plasmon
energies\cite{lundqvist1969characteristic}. This demonstrates that
theories containing additional correlation effects beyond GW theory
can give rise to qualitatively different predictions for the
satellites.

The first-principles GW plus cumulant (GW+C)
approach\cite{lischner2013physical} we use in the present study is a
means to generalize the exact solution of the core electron problem to
the case of dispersing valence
electrons\cite{Hedin,HedinAryasetiawan}. It retains the accuracy of
the first-principles GW approach for quasiparticle properties, but
includes approximately an infinite number of higher order diagrams,
which are needed for an accurate description of satellite properties.

Figures 1(b) and (c) show the calculated photoemission spectra from
the GW plus cumulant (GW+C) and GW approaches, respectively, for the
$\phi=-30^{\circ}$ direction. Both theories predict prominent, intense, occupied
quasiparticle bands at binding energies smaller than 13 eV and a less
intense satellite band structure at higher binding
energies. While the satellite band structures obtained from the GW and
GW+C methods look qualitatively similar, there are several significant
differences: (i) the binding energy of the satellite bands is
significantly larger in the GW method extending to more than 35 eV,
while the GW plus cumulant satellite bands only extend to less than 30
eV, (ii) the total width of the satellite band manifold is 14.4 eV in
the GW approach, significantly larger than the GW plus cumulant theory
width of 10.8 eV and also the quasiparticle band width of 11.7 eV, and
(iii) the distribution of spectral weight is different in the two
approaches. In particular, in the GW approach, the
highest-binding-energy satellite band at 35 eV binding energy in the
vicinity of the $\Gamma$-point is very sharp and intense, while the three
degenerate satellite bands at lower binding energy are broader and
less intense. The GW+C approach does not predict such a sharp,
intensive high-binding-energy satellite band.

\begin{figure}
  \includegraphics[width=8.cm]{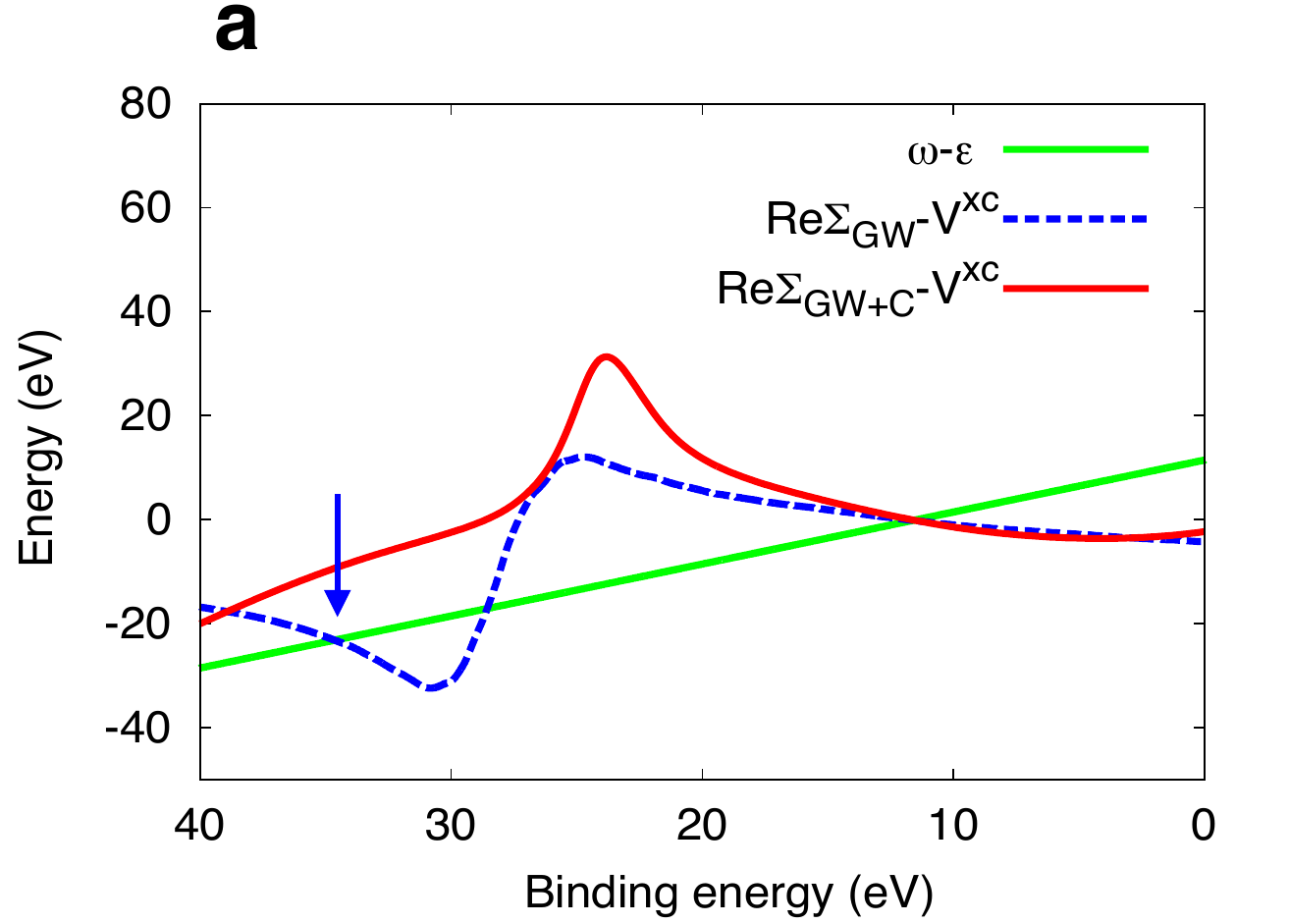}
  \includegraphics[width=8.cm]{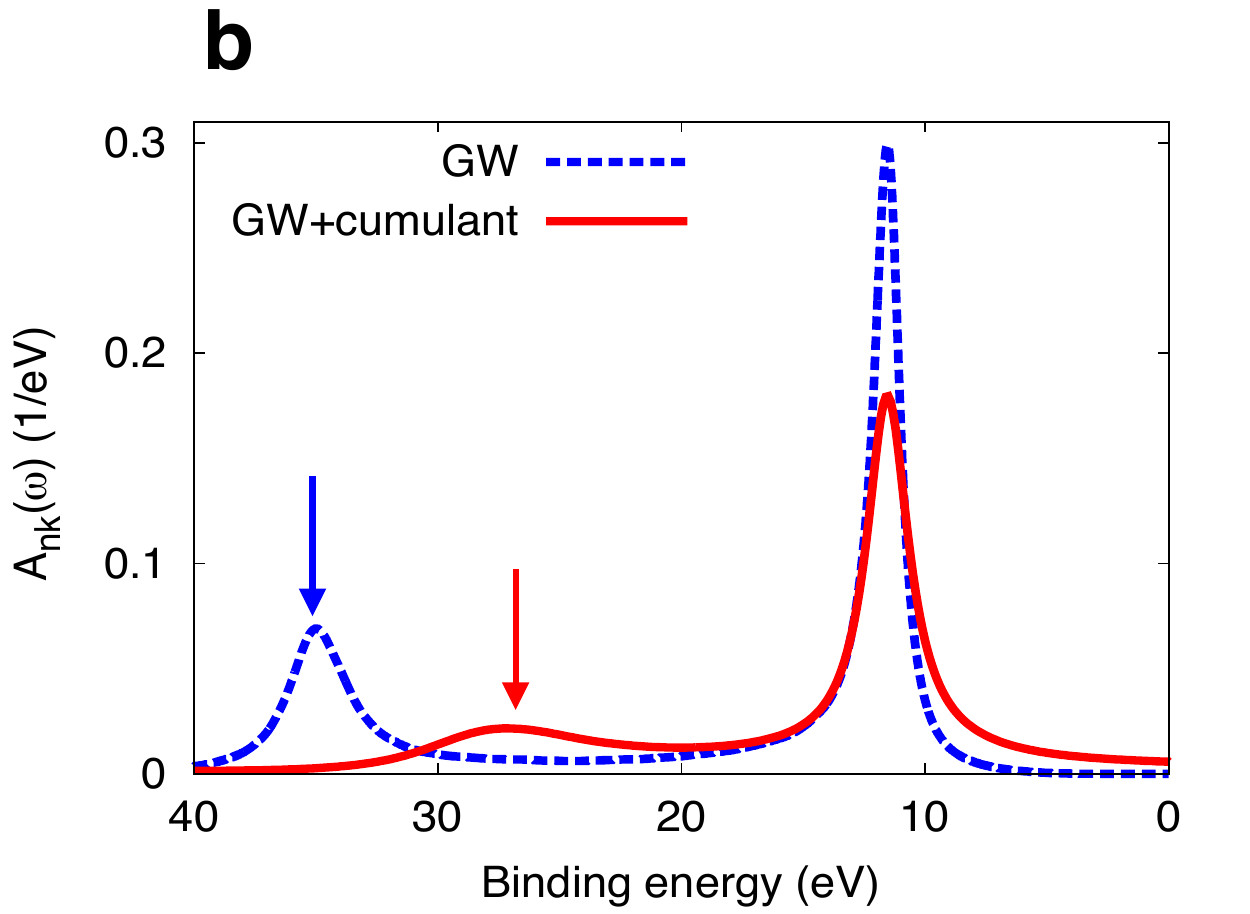}
  \caption{(a): Graphical solution of Dyson's equation for the lowest
    valence band of silicon at the $\Gamma$-point. The blue arrow
    denotes the plasmaron solution of the GW theory. (b): Spectral
    functions for the lowest valence band of silicon at the $\Gamma$-point
    from GW plus cumulant and GW theory. Arrows denote the position of
    the satellite peaks.}
  \label{fig:fig3}
\end{figure}

\emph{Discussion---.} The sharp satellite band at high binding
energies in the GW theory arises from a plasmaron
excitation. Mathematically, well-defined excitations result from
solutions of the quasiparticle or Dyson's equation, $\omega -
\epsilon_{n\bm{k}}=\Sigma_{n\bm{k}}(\omega)-V^{xc}_{n\bm{k}}$, where
$\epsilon_{n\bm{k}}$ denotes the energy obtained from a mean-field
calculation, such as a density-functional theory calculation, and
$V^{xc}_{n\bm{k}}$ denotes the corresponding exchange-correlation
potential. Here, $\Sigma_{n\bm{k}}(\omega)$ denotes the self-energy,
which describes the interaction of the quasi-hole with plasmons and
other excitations. Figure 3(a) shows the graphical solution of the
quasiparticle equation for the $\Gamma$-point of the bulk Brillouin
zone of silicon. If the GW approximation is used to calculate the
self-energy\cite{HedinBook,LouieHybertsen}, we find two solutions: one
solution at low binding energy corresponding to a quasiparticle
excitation and a second solution at a binding energy of 35 eV
corresponding to a plasmaron. In contrast, we do not find a second
solution to the Dyson's equation in the GW plus cumulant
theory. Figure 3(b) shows that the spectral function from GW plus
cumulant theory nevertheless has a second peak, which is separated
from the quasiparticle peak by 16 eV. This separation agrees well with
the calculated and experimentally measured plasmon energy in
silicon\cite{Ehrenreich}, indicating that the satellite results from the creation of
weakly interacting, unbound plasmon-hole pairs. In particular, it can
be shown that the matrix-element weighted density of states of
non-interacting hole-plasmon pairs with a particular wave-vector
$\bm{k}$ has a maximum at the sum of the energy of the hole with wave
vector $\bm{k}$ and the zero wave-vector plasmon energy, if both the
hole and the plasmon have parabolic dispersion relations; and
consequently the satellite band is simply a copy of the hole band
shifted by the zero wave-vector plasmon energy. In contrast, the
separation in the GW theory is 24 eV, indicating strong interactions
between the hole and the plasmons within this lower-order
approximation.

Comparing the theoretical spectral functions of the GW and the GW+C
approaches to the experimental angle-resolved photoemission spectra
(Figures 1 and 2), we find good agreement in both kinds of
calculations for the quasiparticle band structure at binding energies
smaller than 13 eV. However, for the satellite band structure, the agreement of
experiment with GW plus cumulant theory is much better than that with
the GW theory. In particular, the experimental spectrum does not show
a sharp plasmaron band as satellite at 35 eV, in stark contrast with
the prediction of GW theory. Also, the binding energy and the
intensities of the measured satellite bands are in good agreement with
the GW plus cumulant approach, indicating that the satellite band
results from weakly interacting plasmon-hole pairs, very much as is
observed in core-level shake-up plasmon
satellites\cite{Langreth,baird1978angular}, but of course with the
addition of wave-vector dispersion. This shows clearly that the
observed satellite structures originate from the shake-up of plasmons
and not from the formation of plasmarons. Taking into account the good
agreement of recent GW+C calculations with spectroscopic measurements
in nanomaterials\cite{lischner2013physical,lischner2014satellite}, we
conclude that the GW+C method provides a unified picture of
electron-plasmon interactions in materials. This work also calls into
question some prior studies in which plasmarons have been invoked as
relevant excitations
\cite{Rotenberg2,Dial,walter2011effective}. Future work should
investigate the importance of higher-order cumulant functions which so
far have only been studied for electron-phonon
interactions\cite{gunnarsson1994corrections}.

\emph{Acknowledgments.---} G.K.P. acknowledges the Swedish Research
Council for financial support. This work was supported by NSF Grant
No. DMR10-1006184 (theoretical analysis and numerical simulations of
photoemission intensities) and by the SciDAC Program on Excited State
Phenomena (computer codes and algorithm developments) and the Theory
Program (GW and GW+C calculations) at the Lawrence Berkeley National
Lab through the Office of Basic Energy Sciences, US Department of
Energy under Contract No. DE-AC02-05CH11231. C.S.F. acknowledges
salary support from the Lawrence Berkeley National Lab. The
Synchrotron SOLEIL is supported by the Centre National de la Recherche
Scientifique (CNRS) and the Commissariat a l’Energie Atomique et aux
Energies Alternatives (CEA), France. Computer time was provided by the
DOE National Energy Research Scientific Computing Center (NERSC) and
NSF through XSEDE resources at NICS.

\appendix
\section{Appendix}
\emph{Experimental and computational methods.---} As a substrate, we
used a silicon wafer sufficiently conducting (n-doped, 10-20
$\Omega\cdot$cm) in order to avoid charging effects in the
photoemission experiments. The single crystals were cut
($\pm$0.05$^{\circ}$) and polished by Siltronix, with the surface
oriented perpendicular to the [111]-direction. The sample was
introduced into an UHV chamber at a base pressure of $\leq$ 1
$\times$ 10$^{-11}$ mbar and degassed at T = 650 C for 24 hours. The
crystal was then repeatedly flash-heated up to T = 1373 C for a few
seconds by direct current heating. During flash-heating the pressure
remained below p = 5$\times$10$^{-9}$ mbar. This procedure removed the
native oxide layer from the surface and resulted in the equilibrium
structure of Si(111), the well-known 7$\times$7-reconstruction.  This
procedure ensured an atomically flat surface, which is the ideal
starting condition for an ARPES experiment.  To obtain greater bulk
sensitivity and minimize the effects from surface states, a photon
energy of 711 eV was chosen. The photoemission measurements were
performed at liquid nitrogen temperatures to reduce the effects of
thermal diffuse scattering, which led to x-ray photoelectron
diffraction effects superimposed on the measured ARPES spectra. These
effects, although still present in the data, were further separated
out using the procedure in of Bostwick and
coworkers\cite{Rotenberg2}. The experiments were performed at the
ANTARES beam line at the Soleil synchrotron in
Paris\cite{avila2013antares}, France, which employs two X-ray
undulators in tandem, a PGM monochromator combined with a Scienta
R4000 spectrometer. The spectrometer was operated in an angular mode
spanning a 25 or 14 degree angular range with a resolution of 0.1
degrees. The angle between the spectrometer and the photon beam was 45
degrees and all spectra were recorded at normal emission. The
spectrometer resolution was better than 400 meV at pass energy 200 eV
and the photon resolution was 100 meV at h$\nu$ = 711 eV, yielding an
overall instrumental resolution of 130 meV. The binding energy scale
was calibrated using the Au 4f$_{7/2}$ peaks at 84.00 eV of a gold reference
sample.


For the full-frequency GW calculations for silicon, we used the
BerkeleyGW package\cite{BGWpaper}. For the starting mean-field
solution, we carried out density-functional theory (DFT) calculations
within the local density approximation (LDA) using a norm conserving
pseudopotential with a 45 Ry cutoff and an 8$\times$8$\times$8 k-point
grid as implemented in the QUANTUM ESPRESSO program
package\cite{QuantumEspresso}. In the GW calculations, we calculated
the frequency-dependent dielectric matrix in the random phase
approximation (RPA) using 96 empty states and a 5 Ry dielectric
cutoff. We sampled frequencies using a fine grid with a spacing of 0.2
eV up to 150 eV and then a coarser grid up to 300 eV.

\begin{figure}
  \includegraphics[width=8cm]{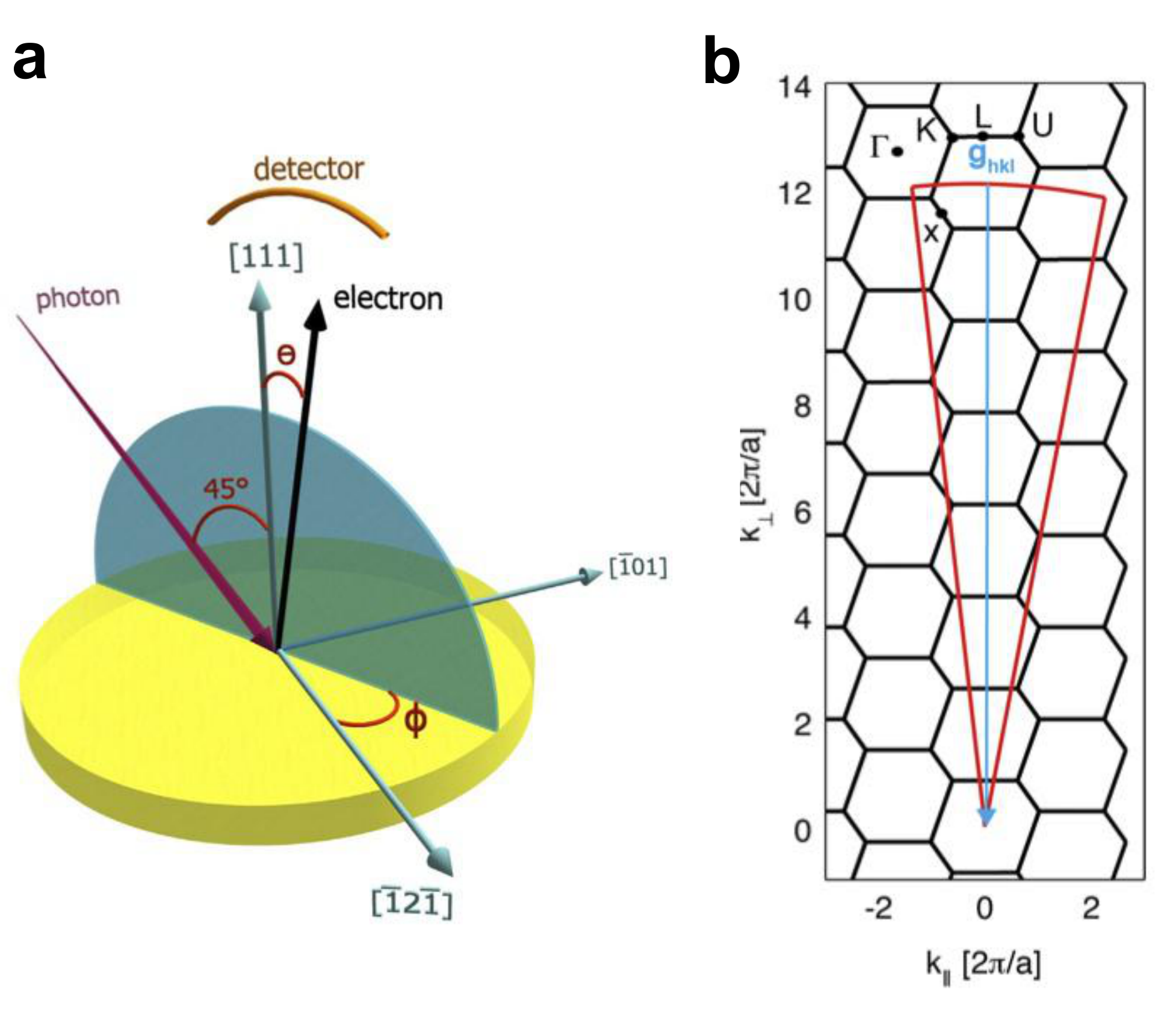}
  \caption{(a): Real space geometry of the photoemission
    measurement. (b): Final-state wave vectors of electrons (red line)
  that reach the detector. The high symmetry points of the Brillouin
  zone of silicon are labeled.}
  \label{fig:App1}
\end{figure}

To describe the final state of the photoelectron, we have employed a
free electron model based upon an inner potential of 12.5 eV, an
average binding energy of 6 eV, and allowance for the work function of
the spectrometer. We have also included effects from the
non-negligible photon momentum. The resulting set of final-state wave
vectors is shown in Fig.~\ref{fig:App1}(b) of the manuscript as the red arc,
which represents the span of the detector in the first Brillouin zones
after translation by the appropriate reciprocal lattice vector
$\bm{G}_{hkl}$. Note that $\bm{k}=0$ in the spectra corresponds to the
$\Gamma$-point of the bulk Brillouin zone, where the three highest
valence bands are degenerate.  Matrix element effects were included by
using tabulated atomic cross sections and projections of the valence
band wave functions onto atomic orbitals.

\emph{First-principles GW plus cumulant theory.---} In the GW plus
cumulant theory\cite{HedinAryasetiawan,Hedin}, the
Green's function for a hole is expressed as
\begin{equation}
G_{n\bm{k}}(t) = i\Theta(-t) \exp\left\{
  -\frac{i\epsilon_{n\bm{k}}t}{\hbar} + C_{n\bm{k}}(t)\right\},
\end{equation}
where $\epsilon_{n\bm{k}}$ denotes the orbital energy from a given mean-field theory
(in this work, a density-functional theory starting point is employed)
and $C_{n\bm{k}}(t)$ denotes the cumulant function. This expression for the
Green's function is obtained after the first iteration of the
self-consistent solution of its equation of motion assuming a simple
quasiparticle form for the starting guess.

The cumulant function can be separated into a quasiparticle part
$C^{qp}_{n\bm{k}}(t)$ and a satellite part $C^{sat}_{n\bm{k}}(t)$ given formally in terms
of the self energy $\Sigma_{n\bm{k}}(\epsilon)$ by (for $t<0$)
\begin{align}
  C^{qp}_{n\bm{k}}(t) &= -\frac{it\Sigma_{n\bm{k}}(E_{n\bm{k}})}{\hbar}
  + \frac{\partial \Sigma^h_{n\bm{k}}(E_{n\bm{k}})}{\partial \epsilon}
  \\
  C^{sat}_{n\bm{k}}(t) &= \frac{1}{\pi} \int_{-\infty}^{\mu} d\epsilon
  \frac{\text{Im}\Sigma_{n\bm{k}}(\epsilon)}{(E_{n\bm{k}}-\epsilon-i\eta)^2} e^{i(E_{n\bm{k}}-\epsilon)t/\hbar},
\end{align} 
where $\mu$ denotes the chemical potential, $\eta$ is a positive
infinitesimal,
$E_{n\bm{k}}=\epsilon_{n\bm{k}}+\Sigma_{n\bm{k}}(E_{n\bm{k}})-V^{xc}_{n\bm{k}}$
is the quasiparticle energy, and $\Sigma_{n\bm{k}}(\epsilon)$ is
defined through the relation
\begin{align}
  \Sigma^h_{n\bm{k}}(\epsilon) = \frac{1}{\pi} \int_{-\infty}^{\mu}
  d\epsilon' \frac{\text{Im}\Sigma_{n\bm{k}}(\epsilon')}{\epsilon'-\epsilon-i\eta}.
\end{align}
For a given level of approximation of $\Sigma$, the cumulant theory
yields an improved Green's function through the above equations. In
this work, we employ the first-principles GW
approximation\cite{LouieHybertsen,hedin1967new} for the self energy,
which gives accurate quasiparticle properties for a wide range of
weakly and moderately correlated semiconductors and insulators.

Having calculated the GW plus cumulant Green's function from the above
set of equations, we obtain the corresponding self energy by inverting
the Dyson equation
\begin{align}
  \Sigma^{GW+C}_{n\bm{k}}(\epsilon)-V^{xc}_{n\bm{k}} =
  \epsilon-\epsilon_{n\bm{k}} + i\eta - G^{-1}_{n\bm{k}}(\epsilon).
\end{align}

\begin{figure}
  \includegraphics[width=9cm]{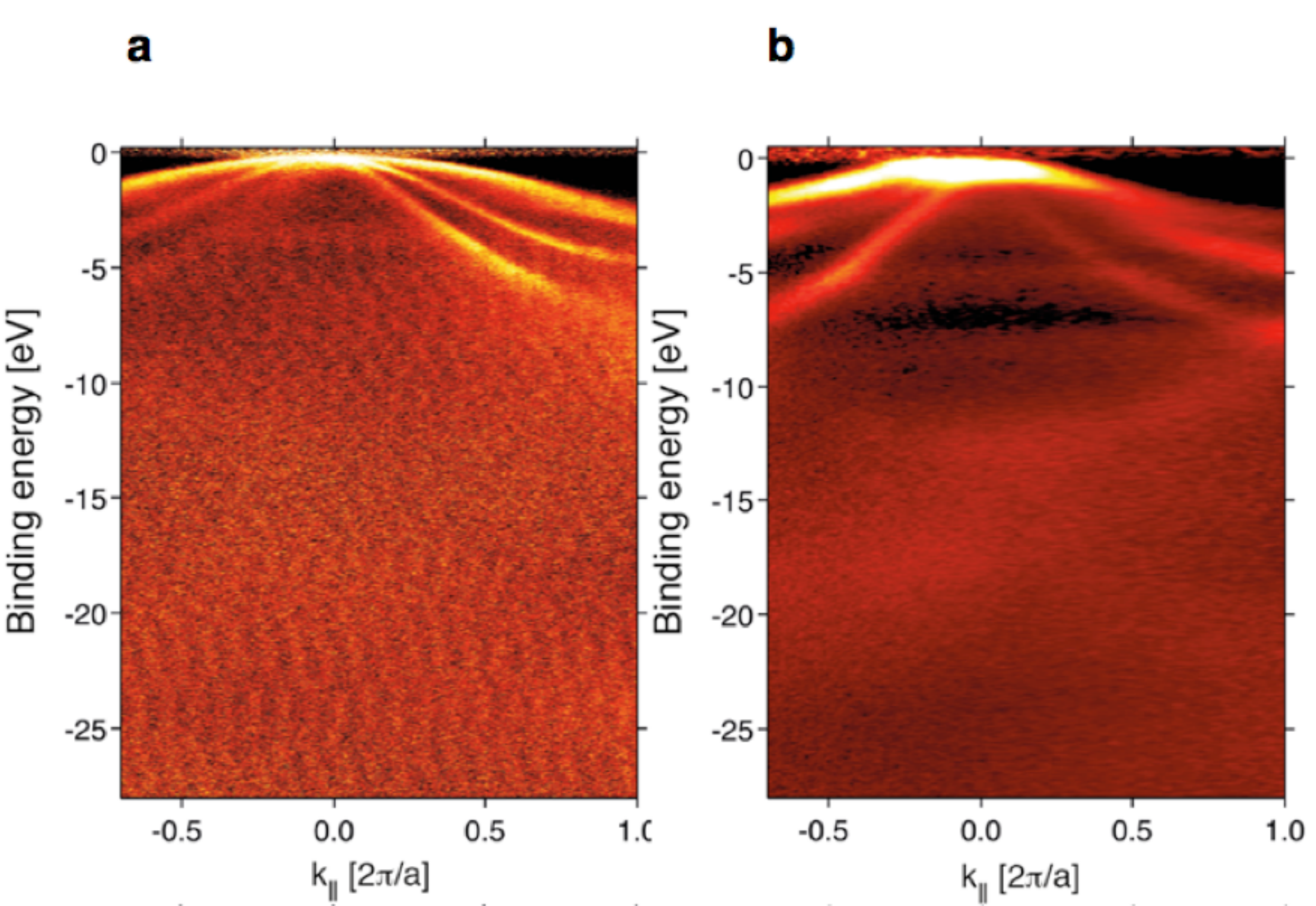}
  \caption{Comparison of the angle-resolved photoemission spectrum of
    silicon at different photon energies. Photon energies of 129 eV
    (a) and 711 eV (b) were used.}
  \label{fig:App2}
\end{figure}

\emph{Angle-resolved photoemission spectrum at 129 eV photon
  energy.---} We have also measured the angle-resolved photoemission
spectrum of silicon at a photon energy of 129 eV. Fig.~\ref{fig:App2}
compares the resulting measured spectrum with the spectrum obtained
with 711 eV photons. Although still present at the lower photon
energy, the plasmon satellite features are much weaker - a consequence
of the reduced extrinsic plasmon losses.

\bibliography{paper}
\end{document}